\documentclass[a4paper]{aa}
\usepackage{psfig,times}

\begin{document}

\newif\ifAMStwofonts
\def\ni{\noindent} 
\def\ea{{et\thinspace al.}\ }                       
\def\eg{{e.g.}\ }                                
\def\ie{{i.e.}\ }                               
\def\cf{{cf.}\ } 
\def\rot{\mathop{\rm rot}\nolimits}
\def\div{\mathop{\rm div}\nolimits} 
\renewcommand{\vec}[1]{\mbox{\boldmath $#1$}}
 
\def\solar{\ifmode_{\mathord\odot}\else$_{\mathord\odot}$\fi} 
\def\gsim{\lower.4ex\hbox{$\;\buildrel >\over{\scriptstyle\sim}\;$}} 
\def\lsim{\lower.4ex\hbox{$\;\buildrel <\over{\scriptstyle\sim}\;$}} 
\def\Ca{Ca\thinspace {\rm II}} 
\def\~  {$\sim$} 
\def\cl {\centerline} 
\def\rl {\rightline} 
\def\x	{\times} 
 
\def\bl{\par\vskip 12pt\noindent} 
\def\bll{\par\vskip 24pt\noindent} 
\def\blll{\par\vskip 36pt\noindent}
\def\alf{$\alpha$}
\def\L{$\Lambda$}
\def\Om{\it \Omega}
\def\nT{$\nu_{\rm T}$\ }
\def\mT{$\mu_{\rm T}$\ }
\def\cT{$\chi_{\rm T}$\ }

\def\apj{{ApJ}}       
\def\apjs{{ Ap. J. Suppl.}} 
\def\apjl{{ Ap. J. Letters}} 
\def\pasp{{ Pub. A.S.P.}} 
\def\mn{{MNRAS}} 
\def\aa{{A\&A}} 
\def\aasup{{ Astr. Ap. Suppl.}} 
\def\baas{{ Bull. A.A.S.}\ } 
\def\csss{{Cool Stars, Stellar Systems, and the Sun}}
\def\an{{Astron. Nachr.}}
\def\sp{{Solar Phys.}}   
\def\gafd{{Geophys. Astrophys. Fluid Dyn.}} 
\def\acta{{Acta Astron.}}
\def\jfm{{J. Fluid Mech.}}

\def\AIP{Astrophysikalisches Institut Potsdam}
\def\F{Ferri\`{e}re}
\def\R{R\"udiger}

\def\qq{\qquad\qquad}                      
\def\qqq{\qquad\qquad\qquad}               
\def\q{\qquad}
\def\bib{\item{}}
\def\top{\item}
\def\toptop{\itemitem}
\def\start{\begin{itemize}}
\def\stop{\end{itemize}}
\def\beg{\begin{equation}}
\def\ende{\end{equation}}
\def\ov{\bar}
\def\om{\omega}
\def\la{\langle}
\def\ra{\rangle}


\title{Do spherical  $\vec{\alpha}^2$-dynamos oscillate? }
\author{ G. \R\inst{1} \and D. Elstner\inst{1} \and  M. Ossendrijver\inst{2}}
\offprints{gruediger@aip.de}
\institute{\AIP,  An der Sternwarte 16, D-14482 Potsdam, Germany \and
 Kiepenheuer-Institut f\"ur Sonnenphysik, Sch\"oneckstr. 6, 79104 Freiburg,
Germany}

\date{\today}

\abstract{The question is answered whether
$\alpha^2$-shell-dynamos are able to produce a cyclic activity or not. Only  kinematic dynamos are considered and  only the solutions with the 
lowest dynamo number are studied without 
restrictions  about the axial symmetry of the solution. The
\alf-effect is allowed to be latitudinally inhomogeneous and/or  
anisotropic, but it is assumed as radially uniform in the turbulent shell.\\
 For a symmetric  \alf-tensor we only find oscillatory solutions if three conditions are simultaneously fulfilled: i) the $\alpha_{zz}$ vanishes or is of the opposite sign as
$\alpha_{\phi\phi}$, ii) the \alf-effect is strongly
concentrated to the equatorial region (i.e. it vanishes at the poles) 
 and iii) the \alf-effect is 
concentrated to a  rather thin outer shell. In the  other cases almost always the
nonaxisymmetric field mode S1 possesses  the lowest dynamo number which
slowly drifts along the azimuthal direction.  Also uniform but 
anisotropic \alf-effect ($\alpha_{zz} = 0$) 
leads to the nonaxisymmetric solutions  as it is  confirmed by the Karlsruhe dynamo experiment. \\ 
However, one of the antisymmetric parts of the
\alf-tensor ({\em not} the vertical magnetic pumping) basically  plays the role of a
differential rotation in the induction equation. Using for the radial
profile of this effect the results of a
numerical simulation for the \alf-tensor of the solar convection zone 
(Ossendrijver et al. 2002), one
indeed finds the possibility of oscillating $\alpha^2$-dynamos even without
the existence of real  nonuniform plasma rotation so that they really can 
be called as  pseudo-\alf$\Om$-dynamos. The resulting butterfly diagram, 
however, proves to be  of the antisolar type. The radial gradient of this
pseudo-differential rotation determines the sign of the phase relation of
$B_r$ and $B_\phi$ (see Sect. \ref{sec4}).
}
\authorrunning{G. \R, D. Elstner \&  M. Ossendrijver}
\titlerunning{Do spherical  $\alpha^2$-dynamos oscillate?  }
\maketitle

\section{Introduction}

\begin{figure}
\hbox{\hspace{1cm}
\psfig{figure=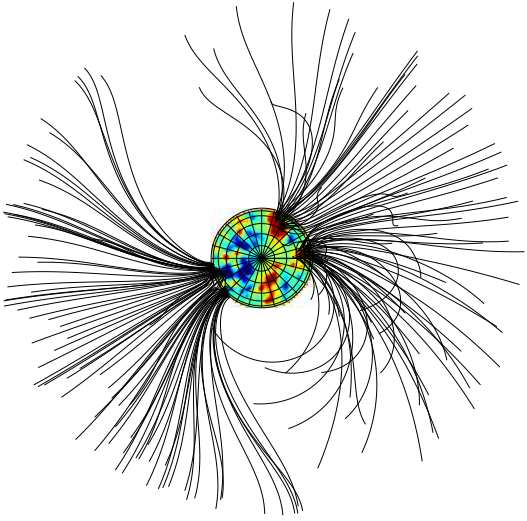,width=6cm,height=8cm}
}
\caption{Pole-on representation of the  magnetic field of AB Dor. Courtesy 
 M. Jardine.}
\label{fff1}
\end{figure}

There are two observations which in the past led to a increasing
interest in the solutions of the relatively simple $\alpha^2$-dynamo. The first
one is the cyclic orbital modulation of close binary systems such as reported by
Hall (1990) and Lanza \& Rodon\`{o} (1999) for RSCVn stars and the second 
one is the flip-flop phenomenon as reported recently by Tuominen et al. 
(1999) and Korhonen et al.
(2001) for FK Coma stars but also in the single young dwarf LQ Hya (K2V, $P_{\rm rot} = 1.6$ days, see
Rice \& Strassmeier 1998). Together with the highly nonaxisymmetric field
configurations for very young cool dwarf  stars reported by Jardine et al.
(2002)  
(see Fig. \ref{fff1}) there is increasing
evidence that the traditional stationary and axisymmetric dipole-solution which
is known since decades does not form the final truth.

Oscillating $\alpha^2$-dynamos have been found to occur in the special
circumstance wherein the $\alpha$-effect changes rapidly in boundary layers 
(R\"adler \& Br\"auer 1987; Baryshnikova \& Shukurov 1987).
In such cases, the period of the $\alpha^2$-dynamo depends strongly 
on the location of the $\alpha$-boundary layer and is typically an
order of magnitude or more smaller than the magnetic diffusion time
across the dynamo-generation shell. In general, oscillatory dynamo
behavior is produced by a combination of both the $\alpha$-effect 
and the $\Om$-effect. Kinematic models of the oscillatory solar dynamo
are always of the $\alpha\Om$-type. 
Here, also motivated by the publication of Schubert \& Zhang (2000), we
discuss the question whether also $\alpha^2$-dynamos alone are able to
exhibit oscillatory solutions. Schubert \& Zhang (2000) considered
$\alpha^2$-dynamos with the (academic) assumptions of completely
homogeneous and isotropic \alf-effect and found oscillatory solutions for
thin outer shells. In Section \ref{3.2} we shall rediscuss this constellation under the extra
condition that only the lowest eigenvalue provides a stable solution. Then
all solutions prove to be non-oscillating.

The present paper only concerns kinematic dynamos on the basis of
\alf-tensors derived by numerical simulations of MHD-convection.
Nonlinearities in the mean-field dynamo equations are not considered. They
are highly complicated as the majority of the resulting magnetic field
configurations is nonaxisymmetric (see Moss \& Brandenburg 1995; K\"uker \&
R\"udiger 1999).

Within the considered turbulent shell the $\alpha$-effect is assumed as
uniform in radius, there is no change of its sign. Stefani \& Gerbeth
(2002) have shown that magnetic oscillations can occur if two shells with
different signs of the $\alpha$-effect exist.\footnote{The oscillating
solution, however, seems to disappear for geometries where one sign of the $\alpha$-effect
dominates the other as it is the case in Fig. \ref{ossen1}.} We do not follow
this possibility in the present paper, here the $\alpha$-effect has been
considered  as uniform in radius within the turbulent shell.
\section{Basic equations and the model}
The model consists of a turbulent fluid in a spherical shell of inner radius
$r_{\rm in}$ and outer radius $r_{\rm out}$. 
A magnetic field is generated in the 
shell by the $\alpha$-effect (Steenbeck \& Krause 1966; Roberts 1972). 
In the shell, the turbulent magnetic diffusivity $\eta_0$ is constant.
For $r>r_{\rm out}$, we assume that there is a conductor with large 
magnetic diffusivity $\eta_{\rm out}$; for $r<r_{\rm in}$,
we assume that there
is a conductor with high electrical conductivity, i.e. small magnetic diffusivity $\eta_{\rm in}$. 
The induction equation is
\begin{equation}
{\partial \vec{B} \over \partial t} = {\rm curl}(\alpha\circ \vec{B}
-{\eta}_{\rm T} {\rm {curl}}{\vec{B}}), 
\label{1}
\end{equation}
where $\vec{B}$ is the magnetic field, $\eta_{\rm T}$ is the turbulent magnetic diffusivity and \alf\ is the
\alf-tensor. In no case the
\alf-tensor has a too simple structure. For astrophysical applications it represents the interaction of an
anisotropic turbulence with a global rotation and a uniform magnetic field.
There are also attempts to include the influence of a nonuniform rotation but
this is beyond the scope of the present study. Here we are only interested in
the structure of the solution of a pure $\alpha^2$-dynamo, i.e. the rotation is
{\em assumed} to be uniform. 

In  R\"udiger  \& Kitchatinov (1993) one finds the overall
structure of the \alf-tensor as 
\begin{eqnarray}
\lefteqn{\alpha_{im}=
 - \alpha_1 (\vec{G}^0 \vec{\Om}^0) \delta_{im} - 
\alpha_2 
(G_i^0 {\Om}_m^0 +
{\Om}_i^0 G_m^0) +}\nonumber\\
\lefteqn{+ \alpha_3(G_m^0 {\Om}_i^0 - G_i^0 {\Om}_m^0)
- \alpha_4(\vec{G}^0 \vec{\Om}^0) {\Om}_i^0
{\Om}_m^0 - \gamma \epsilon_{imk} G_k^0.}
\label{alfim}
\end{eqnarray}
Here the unit vector $\vec \Om^0$ denotes the direction of the axis of the  global rotation of the 
turbulence and the radial unit vector $\vec G^0$ denotes its anisotropy.
In almost all papers about \alf-effect dynamos the expression
(\ref{alfim}) is reduced to its first term of the tensorial 
expression. We
shall demonstrate in the present paper that only the inclusion of the remaining
parts of the \alf-tensor reveals the variety of the solutions of the
$\alpha^2$-dynamo and also solves the problem whether $\alpha^2$-dynamos
can oscillate or not. In the whole paper the influence of the large-scale
flow pattern is ignored. Then the remaining dimensionless number may
be
the dynamo number, $C_\alpha = |\alpha_1|R/\eta_{\rm T}$ with $R$ as the
stellar radius.

It is interesting to consider the antisymmetric parts in the tensor
(\ref{alfim}). A comparison with the induction term ${\cal E}_i= \epsilon_{ikm}
u_k B_m $ reveals that $\gamma_k$ in the last term of (\ref{alfim}) plays the role of
a radial advection (``pumping'') of the magnetic field. On the 
other hand, if formally
a basic rotation with $\vec{u} = \vec{\Om} \times \vec{x}$ is used for the velocity field then one
obtains ${\cal E}_i= (\Om_m x_i - \Om_i x_m) B_m$. By comparison with (\ref{alfim})
it follows that the $\alpha_3$-term in (\ref{alfim}) exactly plays the role of a global
(differential) rotation. More exactly speaking, in cylindrical coordinates ($s,
\phi, z$) we find ($\alpha_{sz} - \alpha_{zs})/2s$ playing the role of an
angular velocity, the gradient of which induces magnetic fields. In Section
\ref{sec4} 
we shall present the influence of the antisymmetric term $\alpha_3$ in (\ref{alfim}) for
the question of the existence of oscillating $\alpha^2$-dynamos.    

\section{The symmetric $\vec{\alpha}$-tensor}
We start in cylindrical coordinates with the symmetric part of the \alf-tensor, i.e.
\beg
\alpha \sim \left( \matrix{-\alpha_1 \cos\theta & 0 & -\alpha_2 \sin\theta\cr
0 & -\alpha_1 \cos\theta & 0 \cr
-\alpha_2 \sin\theta & 0 & -(\alpha_1 + 2\alpha_2 + \alpha_4)
\cos\theta\cr}\right).
\label{alsim}
\ende
As we shall demonstrate, the ratio
\beg
\hat \alpha_z = {\alpha_1 + 2\alpha_2 + \alpha_4 \over \alpha_1}
\label{alfz}
\ende
will be of particular relevance for the resulting solutions. In spherical coordinates ($r, \theta, \phi$) we
have the general structure
\beg
\alpha \sim \left(\matrix{\alpha_{rr} & \alpha_{r \theta} & 0 \cr
\alpha_{\theta r} & \alpha_{\theta\theta} & 0 \cr
0 & 0 & \alpha_{\phi\phi}\cr} \right)
\label{alphsim}
\ende
with $\alpha_{rr} = -(\alpha_1 + 2\alpha_2 + \alpha_4 \cos^2\theta)
\cos\theta$, $\alpha_{r\theta}= \alpha_{\theta r}= (\alpha_2 + \alpha_4 \cos^2\theta)
\sin\theta$, $\alpha_{\theta\theta}= -(\alpha_1 + \alpha_4 \sin^2\theta)
\cos\theta$ and $\alpha_{\phi\phi}= -\alpha_1 \cos\theta$
so that for the ratio (\ref{alfz}) the expression
\beg
\hat \alpha_z = {\alpha_{rr} ({\rm pole}) \over \alpha_{\phi\phi} ({\rm pole})}
\label{alfzz1}
\ende
results where any of the two poles can be taken.

\subsection{The \alf-tensor elements}

Ossendrijver et al. (2001, 2002) presented simulations for all the 
components of the \alf-tensor in spherical coordinates. Some of the 
results, which are relevant for 
the present discussion, are reported here. The simulations were done in a Cartesian box which is meant 
to represent a section from the lower part of a stellar convection zone, including a convectively stable layer 
underneath it. The Cartesian coordinate frame of the box corresponds to the spherical coordinates 
introduced above, such that the $x$-direction corresponds to the negative $\theta$-direction, the $y$-direction 
to the $\phi$-direction, and the $z$-direction (depth) to the negative 
$r$-direction. All simulations were done at 
the southern hemisphere of the star, and the angle between the vertical (radial) direction and the axis of 
rotation was varied between $0\degr$ (south pole) and $90\degr$ (equator). 
Provided the rotation rate is sufficient to yield an \alf-effect, the depth dependence of 
$\alpha_{\theta\theta}$ and $\alpha_{\phi\phi}$ has a typical shape, namely a negative sign in the bulk of 
the thick convection zone, and a positive sign in the thin overshooting layer. These features are expected for the 
southern hemisphere. The amplitudes of $\alpha_{\theta\theta}$ and $\alpha_{\phi\phi}$ increase with increasing 
angular distance from the equator up to a point close to the south pole, more or less consistent with the commonly 
assumed $\cos\theta$-function. Hence, the \alf-effect does {\em not} vanish
at the poles. For weak rotation, the component $\alpha_{rr}$ has a larger amplitude than the 
other two diagonal components, and it has the opposite sign. If rotation increases beyond a certain 
point, $\alpha_{rr}$ as a function of depth developes multiple sign changes, unlike $\alpha_{\theta\theta}$ and 
$\alpha_{\phi\phi}$, and its amplitude falls behind that of the latter. This is the rotational quenching of 
the vertical alpha effect reported in Ossendrijver et al. (2001).
The symmetric component $\alpha_{r\theta}^{\rm S}=\alpha_{zx}^{\rm S}$ is positive in the bulk of the unstable 
layer, and is generally larger in magnitude than $\alpha_{r\theta}^{\rm A}$; it vanishes at the poles.

\begin{figure}
\vbox{
\psfig{figure=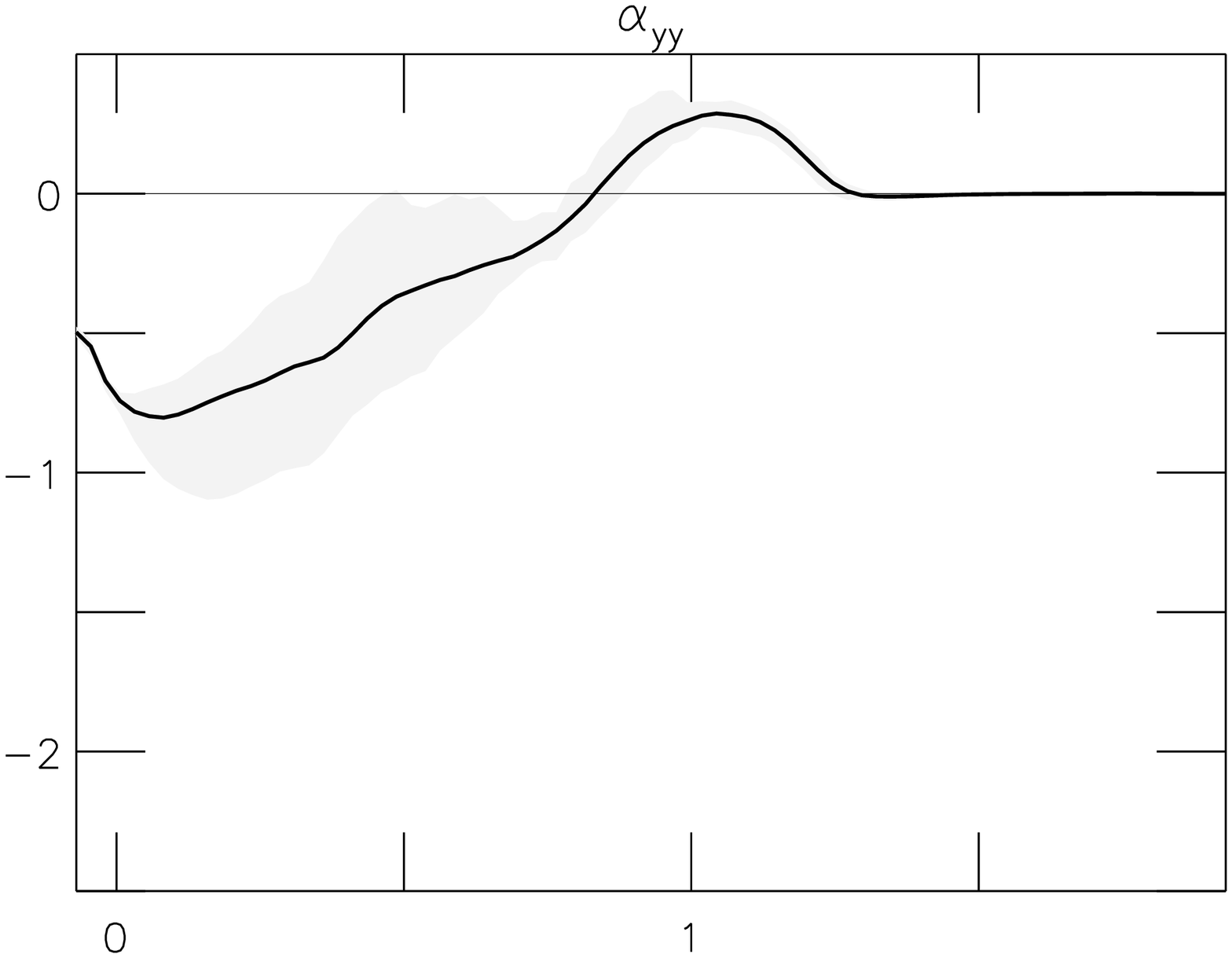,width=8cm,height=3cm}
\psfig{figure=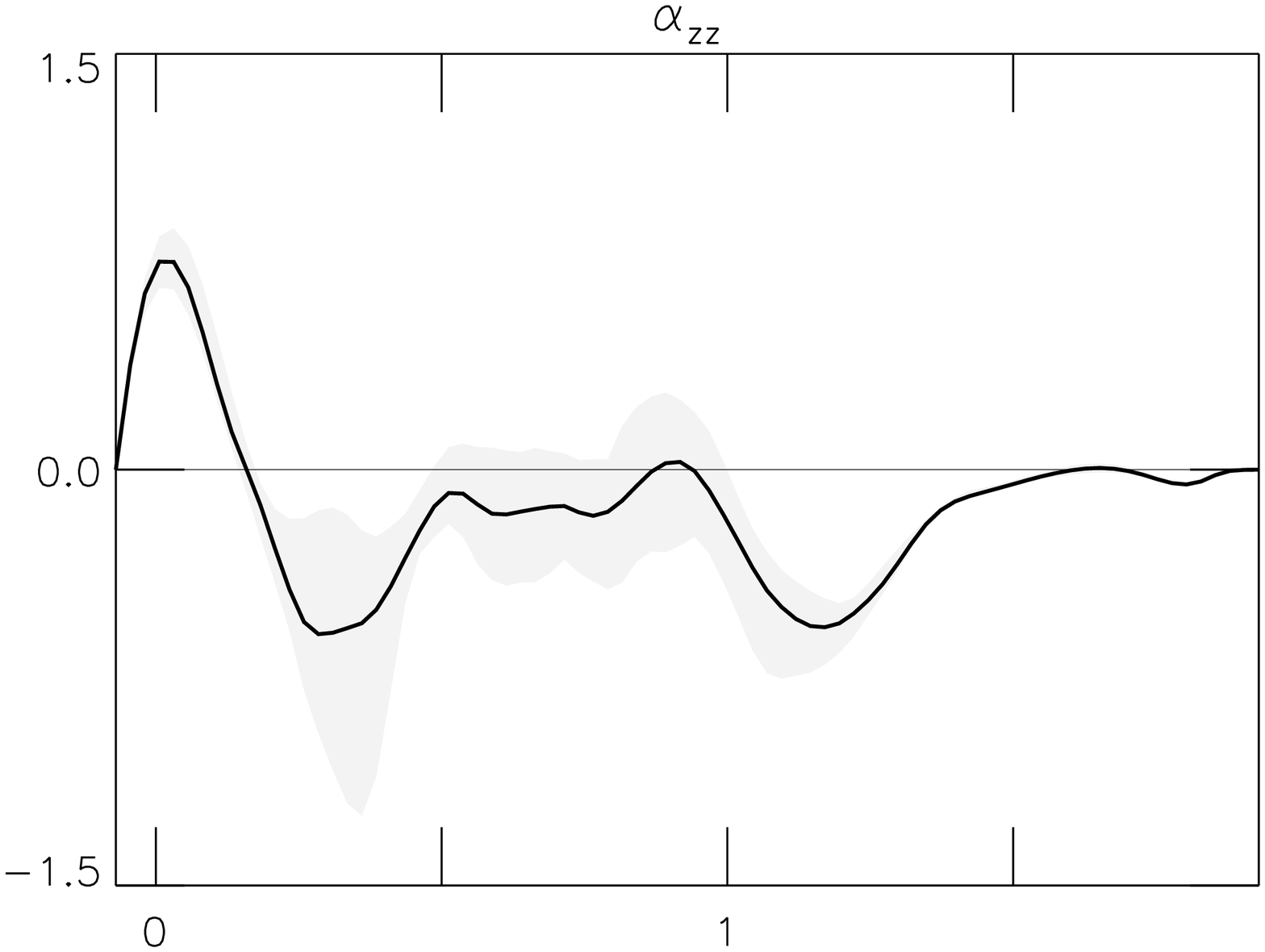,width=8cm,height=3cm}}
\caption{Numerical results for the Cartesian \alf-tensor components $\alpha_{yy}=\alpha_{\phi\phi}$ and $\alpha_{zz}=\alpha_{rr}$, measured in units of 
$0.01\sqrt{dg}$, as a function of depth in units of $d$ from Ossendrijver et al. (2002) for the case of a box 
located at the south pole (run A00). The simulation domain consists of a thin cooling layer ($z<0$), a convectively 
unstable layer ($0<z<d=1$), and a stably stratified layer with overshooting convection ($z>d$). 
The Coriolis number of the run is about $2.4$, which is in the appropriate range for the bottom of the solar
convection zone. For the other parameters and for a detailed description of the model we refer to Ossendrijver 
et al. (2001, 2002). The black curves are spatial and temporal averages; the shaded areas provide an
error indication. Note, that the $\alpha_{\phi\phi}$ does not vanish at the
pole.
}
\label{ossen1}
\end{figure}

\subsection{Results for homogeneous \alf-effect}\label{3.2}

There are references in the literature where the basic antisymmetry of 
the \alf-effect with respect to the equatorial midplane has been neglected 
(R\"adler \& Br\"auer 1987; Schubert \& Zhang 2000; Stefani \& Gerbeth
2000; R\"adler et al. 2002). The only realization of such models can only be
imagined in technical experiments. Indeed, the paper by R\"adler et al. (2002) concerns the
Karlsruhe dynamo experiment with a fixed helicity and a uniform flow field
in vertical ($z$-)direction, so that $\alpha_{zz} = 0$ is obvious. In order
to demonstrate the differences between isotropic and anisotropic
$\alpha$-tensors also in the case of homogeneous \alf-effect (i.e.
$\cos\theta$ ignored), the following calculations are presented for the
cases i) $\alpha_{zz} = \alpha_{\phi\phi}$ and ii) $\alpha_{zz} = 0$. While the
first case seems to be an academic problem, the second case fits the
situation in the dynamo experiment mentioned. 
\begin{figure}
\psfig{figure=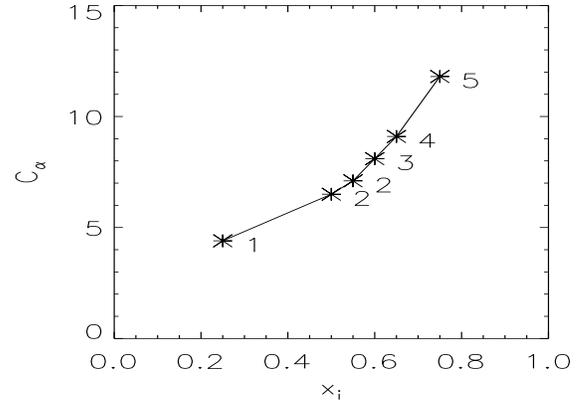,width=9cm,height=6cm}
\caption{Homogeneous and isotropic \alf-effect: The magnetic-field modes 
with the lowest value for $C_\alpha$ for various shell thickness. All the 
solutions  are non-oscillating. The curve is marked with the 
latitudinally mode number $n$ for
which the dynamo number is minimum. The oscillating modes found by Schubert
\& Zhang (2000) are located above the line with the lowest eigenvalues.}
\label{schubert}
\end{figure}

We start
with spherical models embedded in vacuum\footnote{Test computations for
models embedded in a perfect conductor did not reveal basic modifications
of our results in contrast to the  remark by Brandenburg (1994)} with outer \alf-effect zone located between $x=x_{\rm in}$ and $x=1$.
In this zone \alf-effect and eddy diffusivity $\eta_{\rm T}$ are assumed as
radially uniform,
below the convection zone there is a steep transition zone to the
perfect-conducting interior (factor 100). Always the  lowest dynamo 
numbers $C_\alpha = \alpha_1 R/\eta_{\rm T}$ for the 
modes are given in the present paper. Any
mode has an own dynamo number, the mode with the lowest dynamo number is the
preferred mode with the highest stability (Krause \& Meinel 1988).

The 
problem of both homogeneous {\em and} isotropic \alf-effect (i.e.
$\alpha_{im} = \alpha_0 \delta_{im}$ with $\alpha_0 =$ const.) has been
considered by Schubert \& Zhang (2000). Our results do not comply with
theirs if only the solutions with the {\em lowest} dynamo numbers are
considered. In Fig. \ref{schubert} the resulting minimum dynamo numbers
$C_\alpha$ and the associated latitudinal mode number $n$ are given for 
various inner shell 
radius $x_{\rm in}$. In opposition to the case of inhomogeneous
\alf-effect ($\sim \cos\theta$) the solution of the induction equation 
(\ref{1}) are modes with a single latitudinal mode number $n$. All the 
solutions
can thus be characterized by the mode number $n$; the solutions with the
mode number $n$ {\em and} with the lowest eigenvalue $C_\alpha$ are given
in Fig. \ref{schubert} -- they all are non-oscillating.  We do not find 
any oscillating \alf$^2$-dynamo  unless its \alf-amplitude was not the 
lowest one. As we take from the Fig. \ref{schubert}, however, it is clear
that it is {\em not} enough to consider only the modes with $n=1$ or  
$n=2$. The thinner the (outer) \alf-shell the higher is the latitudinal
mode number $n$ of the modes excited with the lowest dynamo number
$C_\alpha$. Oscillating modes with low $n$ are also existing but for thin
shells their
$C_\alpha$ is never the lowest one.\footnote{The same remark concerns the
analysis of R\"adler \& Br\"auer (1987).} 
\begin{table}[t]
\caption{Dynamo numbers  for anisotropic ($\alpha_{zz} = 0$) but uniform 
\alf-effect ($\alpha =$ const., cf. Karlsruhe dynamo experiment). The
boldface numbers mark the magnetic mode with the lowest eigenvalue}
\begin{tabular}{|l||l|l|l|l|}
\hline
&&&&  \\[-1.5ex]
 $x_{\rm in}$& A0 & S0 & A1 & S1 \\[1ex]
  \hline 
 &&&& \\[-1.5ex]
  0.25& 9.06 & 8.40 & {\bf 6.72} & {\bf 6.72}\\[0.5ex]
  0.50& 10.84 (osc) & 10.64 (osc) & {\bf 9.50} & {\bf 9.50}\\[0.5ex]
  0.75& 17.29 (osc) & 17.32 (osc) & {\bf 16.23} & {\bf 16.23}\\
\hline
\end{tabular}
\label{tab00}
\end{table}

\begin{table}[t]
\caption{ The same as in Table \ref{tab00} but for   $\alpha \sim
\cos\theta$  }
\begin{tabular}{|l||l|l|l|l|}
\hline
&&&&  \\[-1.5ex]
 $x_{\rm in}$& A0 & S0 & A1 & S1 \\[1ex]
  \hline 
 &&&& \\[-1.5ex]
  0.25& 14.99 (osc) & 14.94 (osc) & {10.64} & {\bf 9.97}\\[0.5ex]
  0.50& 15.7 (osc) & 15.5 (osc) & {11.8} & {\bf 11.7}\\[0.5ex]
  0.75& 21.58 (osc) & 21.58 (osc) & {\bf 18.43} & {\bf 18.43}\\
\hline
\end{tabular}
\label{tab000}
\end{table}
More interesting are the solutions with homogeneous but anisotropic
\alf-tensor. The \alf-tensor has no $zz$-component, i.e. $\alpha_{zz} = 0$
in cylindric coordinates.  Now the rotation axis is clearly defined so 
that it makes sense to ask for the axisymmetry of the solutions. Table 
\ref{tab00} gives the results for uniform
$\alpha$-effect and Table \ref{tab000} gives the results if the \alf-effect
is antisymmetric with respect to the equator, i.e. 
$\alpha \sim \cos\theta$. It was important to include 
the nonaxisymmetric modes into the consideration as they indeed possess the 
lowest dynamo numbers. Always the nonaxisymmetric
modes with $m=1$  dominate for all the models considered; and there are 
again no oscillations.  The axisymmetric solutions (mostly oscillating) which 
have been found by Busse \& Miin (1979), Weisshaar (1982) and Olson 
\& Hagee (1990) are probably not stable as the solution with the lowest dynamo number are  
nonaxisymmetric and azimuthally drifting rather than axisymmetric and oscillating.

There are no basic differences for the cases given in Tables \ref{tab00} and
\ref{tab000}, i.e. for $\alpha \sim$ const. and $\alpha \sim \cos\theta$
 ($\alpha_{zz} = 0$ in both cases). The dominance of the $m=1$ modes for anisotropic \alf-effect is a well-known 
result which has already been presented by R\"udiger (1980) and by R\"udiger  \& Elstner (1994).  
Here we have added its relevance also for the more simple case 
of  {\em homogeneous} \alf-effect. In the light of these calculations 
it is thus no surprise that the Karlsruhe dynamo experiment indeed 
provides the nonaxisymmetric modes with  $m=1$ (Stieglitz \& M\"uller (2001).

\subsection{Results for inhomogeneous \alf-effect}
In the following we shall develop further the models with the equatorial antisymmetry of the \alf-effect as it results as the consequence of  a global rotation of the considered spherical object. But we shall  consider different $\alpha$-profiles in latitude in order to simulate a possible concentration of the \alf-effect to the equator. We fix  the latitudinal profile such as 
\begin{equation}
\alpha\propto  \sin^{2\lambda}{\theta} \cos\theta,
\end{equation}
where $\lambda$ is a free parameter describing the latitudinal profile of 
the \alf-effect. For $\lambda>0 $ the \alf-effect at the poles vanishes. 
It  is more and more concentrated  at lower latitudes for 
  increasing value of $\lambda$. Note, however, that the box simulations
  did not reveal basic deviations of the latitudinal \alf-profile from the
  $\cos\theta$-law. In particular, at the poles the effect did {\em not} 
  vanish. Insofar, the dynamo models with $\lambda > 0$ seem to be only of
  academic interest, but it is interesting to know that only for such
  models oscillating solutions appear with the lowest eigenvalues.

The  results for kinematic dynamo models with the (realistic) 
equatorial antisymmetry are summarized in the Tables
\ref{tab1}$\dots$\ref{tab3} presenting  the
critical eigenvalues  $C_\alpha$   for different $\alpha$-tensor  models and convection
zones with various depths.  Our notation is the standard one,
i.e. A$m$ denotes a solution with
antisymmetry with respect to the equator and with the azimuthal quantum 
number $m$  (see Krause \& R\"adler 1980). The oscillating solar magnetic field mode  mode is antisymmetric with
respect to the equator, it is of A0 type. 

Table \ref{tab1} gives the results for {\em isotropic \alf-effect} 
($\alpha_{z}=1$). Again the boldface numbers represent the absolutely lowest eigenvalues $C_\alpha$ indicating maximal
stability. For the standard case with $\lambda=0$ Table \ref{tab1} provides the axisymmetric
dipole  A0 as the stable mode. This result, however, strongly
depends on the latitudinal profile of the \alf-effect.  Already for $\lambda=1$ the preferred mode is  nonaxisymmetric, i.e. A1 and/or  S1 and for
$\lambda=2$ and 3 we always find
the  A1 mode as the preferred one.  Our result is that except for the simplest latitudinal profile of the \alf-effect (i.e. the $\cos \theta$-dependence) the solutions are no longer axisymmetric, so that the work with axisymmetric codes only has a very restricted meaning. Although  oscillating modes also appear they
never have the lowest eigenvalue. Note that from all our latitudinal \alf-profiles only the $\cos \theta$-dependence leads to finite \alf-values at the poles.

For anisotropic
\alf-tensor we even have a more clear situation. Simplifying we worked with $\alpha_{zz}=0$. The Table \ref{tab2} 
presents the results. Always the  solutions are nonaxisymmetric as always the oscillating axisymmetric modes possess higher eigenvalues. They occur, however, for the same
\alf-anisotropy but for the  thin convection zones with 
$x_{\rm in} = 0.8$ and $\lambda>0$ 
(Table \ref{tab3}). For such a model where the \alf-effect does {\em not} exist in the polar
regions we    find  that the mode with the lowest
eigenvalue (i.e. the stable mode) forms  oscillating axisymmetric magnetic
fields (of quadrupolar equatorial symmetry, no dipoles).  It should be underlined, 
however, that such a cyclic behavior seems to be a rather exceptional case as it  only appears if three conditions are fulfilled, i.e.
\begin{itemize}
\item[i)] the \alf-tensor must be highly anisotropic,
\item[ii)] the \alf-effect must be concentrated to the equator,
\item[iii)]  the convection zone is rather thin.
\end{itemize}
The latter condition  remembers a similar finding of R\"adler \& Br\"auer 
(1987)  -- but in contrast to their consideration this thin-shell 
condition is   {\em not} a sufficient one for oscillating \alf$^2$-dynamos.

\begin{table}[t]
\caption{Dynamo numbers $C_\alpha$ for isotropic $\alpha$-effect ($\alpha_z
= 1$).  The bottom of the 
convection zone is at $x_{\rm in}=0.5$}
\begin{tabular}{|l||l|l|l|l|}
\hline
&&&&  \\[-1.5ex]
 $\lambda$& A0 & S0 & A1 & S1 \\[1ex]
  \hline 
 &&&& \\[-1.5ex]
  0& {\bf 9.41} & { 9.42} & 9.75 & 9.76 \\[0.5ex]
  1& 28.8 (osc) & 28.7 (osc) & {\bf 26.7} & {\bf 26.7}\\[0.5ex]
  2& 41.1 (osc) & 41.2 (osc) & {\bf 38.8} & 39.2\\[0.5ex]
  3& 51.9 (osc) & 52.0 (osc) & {\bf 49.8} & 50.3\\
\hline
\end{tabular}
\label{tab1}
\end{table}

\begin{table}[t]
\caption{Eigenvalues $C_\alpha$ for $\alpha_{zz}=0$  and  $x_{\rm in}=0.5$}
\begin{tabular}{|l||l|l|l|l|}
\hline
&&&& \\[-1.5ex]
 $\lambda$& A0 & S0 & A1 & S1 \\[1ex]
  \hline
&&&&   \\[-1.5ex]
  0& 15.7 (osc) & 15.5 (osc) & { 11.8} & {\bf 11.7} \\[0.5ex]
  1& 35.0 (osc) & 33.8 (osc) & { 32.7} & {\bf 31.3}\\[0.5ex]
  2& {51.7}(osc)  & 49.1 (osc) & 49.3 & {\bf 47.0}\\[0.5ex]
  3& { 66.6} (osc) & 63.3 (osc) & 64.3 & {\bf 61.4}\\
\hline
\end{tabular}
\label{tab2}
\end{table}

\begin{table}[t]
\caption{Eigenvalues $C_\alpha$ for $\alpha_{zz}=0$  and  $x_{\rm in}=0.8$}
\begin{tabular}{|l||l|l|l|l|}
\hline
&&&& \\[-1.5ex]
 $\lambda$& A0 & S0 & A1 & S1 \\[1ex]
  \hline
&&&&   \\[-1.5ex]
  0& 26.3 (osc) & 26.3 (osc) & {\bf 23.2} & {\bf 23.2} \\[0.5ex]
  1& 63.0 (osc) & {\bf 62.9} (osc) & { 63.0} & {\bf 62.9}\\[0.5ex]
  2& {89.8}(osc)  & {\bf 89.4} (osc) & 90.2 & { 89.9}\\[0.5ex]
  3& { 112.9} (osc) & {\bf 111.8} (osc) & 113.4 & {112.7}\\
\hline
\end{tabular}
\label{tab3}
\end{table}

\section{The antisymmetric $\vec{\alpha}$-tensor}\label{sec4}
Let us now turn to the antisymmetric parts of the \alf-tensor. There is at
first the $\alpha_3$-component in the tensor formulation (\ref{alfim}). As
mentioned above, the formation $(\alpha_{\theta r} -
\alpha_{r\theta})/(2r\sin\theta)$ formally acts as a (differential)
rotation -- so that in reality, if $\alpha_3$ is not too small -- {\em all
$\alpha^2$-dynamos can operate as (pseudo) $\alpha\Om$-dynamos} which are known as
oscillatory. We shall denote this virtual
angular velocity by $\Om_{\rm T}$ with
\beg
{\Om}_{\rm T} = - {\alpha_3 \over r}.
\label{omt}
\ende
Obviously, the ratio $\alpha_3/\alpha_1$ will determine the ability of the
$\alpha^2$-dynamo to operate as a (pseudo) $\alpha\Om$-dynamo. In any case,
however, it transforms poloidal magnetic fields to toroidal magnetic fields
with a phase relation depending on the sign of $\partial \Om_{\rm
T}/\partial r$\footnote{sign $(B_r B_\phi) = $ sign $(\partial \Om_{\rm
T}/\partial r)$}.

In spherical coordinates the antisymmetric parts of the \alf-tensor
(\ref{alfim}) can be written as
\beg
\alpha \sim \left(\matrix{0 & \sin\theta \ \alpha_3 & 0\cr
-\sin\theta \ \alpha_3 & 0 & -\gamma\cr
0 & \gamma & 0\cr}\right).
\label{alphim} 
\ende
With the notation by Ossendrijver et al. (2002) one finds $\alpha_3 =
-\gamma_\phi/\sin\theta$, i.e. $\alpha_3 = -\gamma_\phi$(equ).

\subsection{Simulations}
From the simulations of Ossendrijver et al. (2002) the following numerical results on $\gamma_{\phi}$ and $\gamma_r$
can be mentioned. Longitudinal pumping ($\gamma_{\phi}$) is generally the strongest pumping effect observed in the 
simulations unless the rotation axis is in the radial direction. It has a predominantly negative sign
within the bulk of the unstable layer and the overshoot layer, which signifies that the mean field is advected 
in the retrograde direction. In most cases reported in Ossendrijver et al. (2002), there is also a thin layer near 
the top of the convection zone where the field is pumped in the prograde direction. The longitudinal pumping 
effect is strongly dependent on latitude; it vanishes at the pole and peaks at the equator.

The direction of vertical pumping ($\gamma_r$) is downward ($\gamma_r<0$; i.e. $\gamma_z>0$) in the bulk of 
the unstable layer and in the overshoot layer. Near the top of the box there is a thin layer where the pumping is 
directed upwards ($\gamma_r>0$). There is little dependence on latitude or on rotation.

\begin{figure}
\vbox{\vspace{1.5cm}
\psfig{figure=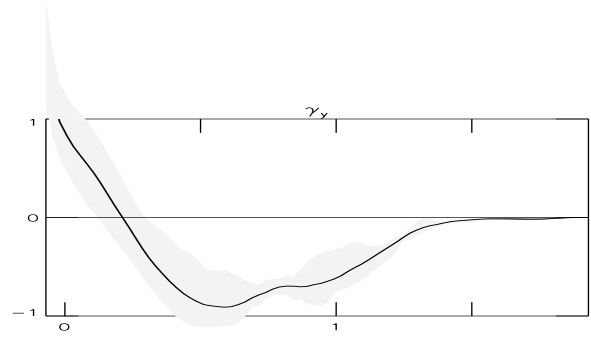,width=8cm,height=3cm}
}
\caption{Pumping velocity $\gamma_{\phi}=\gamma_y$, measured in units of $0.01\sqrt{dg}$, as a function of depth 
in units of $d$ from Ossendrijver et al. (2002) for the case of a box located at the  equator (their Fig. 4).
The tensor quantity $\gamma_\phi$  equals 
$r\Om_{\rm T}$.}
\label{ossen3}
\end{figure}

\begin{figure}
\psfig{figure=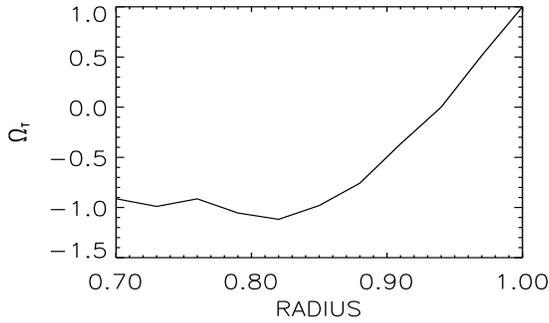,width=8cm,height=5cm}
\caption{The same as in Fig. \ref{ossen1} but in the representation 
${\Om}_{\rm T}={\Om}_{\rm T}(r)$ (see Eq. \ref{omt}).}
\label{ossen2}
\end{figure}
\subsection{Results}
We start with the (unrealistic) case of radially uniform $\alpha_3$. Then after
(\ref{omt}) is $\partial{\Om}/\partial r = \alpha_3/r^2$, so that for 
positive $\alpha_3$ there seems to be superrotation. Characteristic results are given in
Table \ref{tab0}. For too small $\alpha_3$, the nonaxisymmetric solution (A1)
of the $\alpha^2$-dynamo is hardly influenced. For $\alpha_3 \simeq 10$,
however, we already find an oscillating quadrupole. Between $\alpha_3 =5$ and
$\alpha_3 = 10$ there is the transition of the kinematic dynamo from
nonaxisymmetric drifting modes to axisymmetric oscillating modes.

Obviously, the considered values of $\alpha_3$ for the transition to a 
pseudo $\alpha\Om$-dynamo are rather 
high. It is the non-uniformity of the $\Om_{\rm T}$-effect which appears
in the induction equation and may only allow the oscillations. We have thus applied to the model 
the radial profile in Fig. \ref{ossen1} which results from real 
simulations (Ossendrijver et al. 2002). The profile
is multiplied with the amplitude $\hat \alpha_3$ which is varied in order to
find the various dynamo solutions. The results are given in Table
\ref{tab01} and they are a surprise. Indeed, the amplification of the 
$\alpha_3$-effect only by a   factor of 3 leads to the appearance of 
oscillating solutions as the most stable one. Again it shows quadrupolar 
equatorial symmetry. Dipolar symmetry only appears for the amplification 
factor of $-10$, i.e. the (say) $\alpha{\Om}_{\rm T}$-dynamo resulting by the simulations of Ossendrijver et al.  may lead to oscillating solutions but with quadrupolar symmetry.
\begin{table}[t]
\caption{Dynamo numbers for anisotropic \alf\ ($\alpha_{z} = 0$) and  
uniform $\alpha_3$-effect.  $x_{\rm in} =0.5$, $\alpha_2=0$}
\begin{tabular}{|c||c|c|c|c|}
\hline
&& && \\[-1.5ex]
 $\hat \alpha_3$& A0 & S0 & A1 & S1 \\[1ex]
  \hline 
 && &&\\[-1.5ex]
  3 & 14.9 (osc) & 14.4 (osc) & 12.1 & {\bf 12.0} \\[0.5ex]
  5 & 14.3 (osc) & 13.7 (osc) & 12.9 & {\bf 12.9} \\[0.5ex]
  10 & 12.9 (osc) & {\bf 12.0} (osc) & 15.0 & 15.1\\
  \hline
\end{tabular}
\label{tab0}
\end{table}

\begin{table}[t]
\caption{Dynamo numbers for anisotropic \alf\  ($\hat \alpha_{z} = 0$) 
and  nonuniform $\alpha_3$-effect (see Figs. \ref{ossen1}, \ref{ossen2}).  
$x_{\rm in} =0.5$, $\alpha_2=0$  }
\begin{tabular}{|r||c|c|c|c|}
\hline
&& && \\[-1.5ex]
 $\hat\alpha_3$& A0 & S0 & A1 & S1 \\[1ex]
  \hline 
 && &&\\[-1.5ex]
  1 & 15.1 (osc) & 14.8 (osc) & 12.1 & {\bf 11.9} \\[0.5ex]
   2 & 14.9 (osc) & 14.5 (osc) & 12.9 & {\bf 12.8} \\[0.5ex]
  3 & 15.0 (osc) & {\bf 14.6}(osc) & 17.4 & 17.4 \\[0.5ex]
  \hline
  $-10$ & {\bf 14.7} (osc) & {\bf 14.7} (osc) & {16.6} & {16.6}\\
  \hline
\end{tabular}
\label{tab01}
\end{table}

\section{Discussion}
We have shown that for spherical configurations with outer turbulent shells
the basic solution for $\alpha^2$-dynamos is a nonaxisymmetric mode
drifting in the azimuthal direction. Oscillating axisymmetric solutions are
seldom exceptions for \alf-effects that are
\begin{itemize}
\item[i)] highly anisotropic,
\item[ii)] strongly concentrated to the equatorial region, 
\item[iii)] restricted to thin outer shells. 
\end{itemize}
In all other cases the mode with
the lowest eigenvalue (which are considered here as the only stable one) is
nonaxisymmetric and almost always of quadrupolar symmetry. The azimuthal
drift is always of the same order as the oscillation frequencies are. Note that the box simulations by
Ossendrijver et al. (2001) did {\em not} lead to vanishing \alf-effect at the poles. 

The same is true for $\alpha^2$-dynamos under inclusion of those
anisotropic parts of the \alf-tensor which can be combined to 
antisymmetric components, i.e. $G_i {\Om}_m - G_m {\Om}_i$. This term
corresponds to a (pseudo-) angular velocity, see Eq. (\ref{omt}), which induces electrical fields in the same way as a
real differential rotation of the plasma would do. The box simulations of
Ossendrijver et al. reveal this pseudo-${\Om}$ (called here ${\Om}_{\rm
T}$) as increasing outwards (see Fig. \ref{ossen2}) so that in any case,
together with the positive $\alpha_{\phi\phi}$ (in the northern
hemisphere), 
an antisolar butterfly diagram results if the solution was axisymmetric
and oscillating. The necessary dipolar solution, however,  only exists if the ${\Om}_{\rm T}$-effect is
artifically amplified by a factor of 10 (see Table \ref{tab01}). 

We conclude that shell-dynamos without differential rotation on the 
basis of MHD simulations  of rotating stellar convection always possess 
nonaxisymmetric magnetic field configurations such as recently found for 
AB Dor.  Oscillating solutions for \alf$^2$-dynamos are revealed as rather exceptions. Our conclusion is that a cyclic stellar activity can always be considered as a strong indication for the existence of internal differential rotation.

\end{document}